# Determination of fast electron energy dependence of magic angles in electron energy loss spectroscopy for anisotropic systems

Y.K. Sun and J. Yuan

Department of Materials Science and Engineering, Tsinghua University, Beijing,

100084, P.R. China

**Abstract**

We present an accurate determination of the magic angle conditions at which the sample-orientation induced fine structure variation is eliminated in the core-level electron energy loss spectroscopy of anisotropic systems.  Our result paves the way for the application of magic angle electron energy loss spectroscopy (MAEELS) in material characterization.  It also highlights, for the first time, the importance of the quantum interference effect between longitudinal and transverse interactions for medium energy fast electrons, and its connection with the anisotropy of the electron transition.

Inelastic scattering of fast charged particle is a well-known physical phenomenon[1,2] and has been used to study the internal structure of matters from fundamental particles to atoms, molecules, clusters, as well as condensed matters. For example, the unique ability to focus a high-energy electron beam down to the sub-nanometer scales allows the electron energy loss spectroscopy (EELS) to be a powerful tool to study the electronic structures of materials and devices, many of which are intrinsically anisotropic[3,4] or potentially anisotropic because of symmetry-breaking due to either deformation or close proximity to surfaces or boundaries[5]. Measuring anisotropic electronic structures is complicated by the sample-orientation dependence of the transition matrix element[6]. The similar problem is also known in other forms of spectroscopy. For example, the anisotropy induced complexity in nuclear magnetic resonance (NMR) of solid state materials is removed with the application of the magic angle spinning (MAS) technique[7]. The quantitative study of electronic excitation for anisotropic materials would benefit from a similar technique in EELS. Menon and Yuan[8] have shown that if we illuminate the anisotropic sample with a parallel beam and detect the inelastic scattered electrons with a circular axial detector characterized by a collection semi-angle $\theta_0$ (Fig.1a), a 'magic' semi-angular (MA) $\theta_{magic}$ can be found to offer orientation independent EELS spectra. This technique is called Magic Angle Electron Energy Loss Spectroscopy (MAEELS) which can be generalized to convergence beams[9]. As the spectra collected is equivalent to spherically averaged one that can be interpreted in a straightforward manner without the need for the sample-orientation information[9], MAEELS is very important for spatially resolved quantitative investigation of anisotropic electronic structures of, for example, defect structures in carbon nanotubes[10] and dislocation core in light emitting wide band-gap semiconductors[3]. However, the wide spread use of MAEELS is hampered by the lack of the precise MA condition whose determination is both difficult experimentally and controversial theoretically[11-13]. In this Letter, we present a careful experimental determination of MA as a function of both the fast electron energy $E_0$ and the energy loss $\Delta E$ involved. The discrepancy of the experiment with the existing inelastic scattering theories has been account for by relativistic effects. By recognizing the importance of quantum interference phenomenon, a subject of fundamental interest on its own[14], and modifying the traditional incoherent approach[1] of relativistic scattering theory, we have reached a quantitative model that is in excellent agreement with our refined experimental results.

The experimental geometry to determine MA conditions is shown in Fig.1a. Using a well-known uniaxial system of highly oriented pyrolytic graphite (HOPG) as an example, Fig.1b shows a series of EELS spectra of carbon 1s absorption. The experiment was performed in the diffraction couple mode of a JEM-2010 field emission transmission electron microscope with a Gatan Imaging Filtering prism-type energy-loss spectrometer and the incident electron accelerated to 200keV. The spectral fine structure shows marked variation as a function of both sample orientation (defined by the angle γ between the graphite c-axis and the incident beam axis) and the collection semi-angle $\theta_0$ (Fig.1a). The simplest way to approximately determine MA is to compare the spectra for similarity at all sample orientations, and an example is shown in (ii) and (v) of Fig.1(b). A quantitative method is to measure the normalized intensity variation of anisotropic transition against the collection semi-angle[13]. For carbon 1s spectrum of graphite, there are two main anisotropic contributions from 1s->2p($\pi^*$) (shown as the pre-peak at 285eV) and 1s->2p($\sigma^*$) (shown as a broad peak at 292eV). The ratio R of $I_{\pi^*}$ of the pre-peak intensity to total intensity ($I_{\pi^*}+I_{\sigma^*}$) was deduced from each spectrum after subtraction of the pre-edge background and careful deconvolution of the multiple scattering effect. The experiments were repeated for a systematic variation of $\theta_0$ and the specimen tilt (varying $\gamma$). For each specimen tilt series, the data was fitted to the following formula which should describe the orientation dependence of the signal[9]:

$$R(\gamma,\theta_0)=R_{MA}+F_c(\theta_0)\cdot(\cos^2\gamma-1/3) \qquad (1)$$

where $R_{MA}$ is a constant, $F_c$ is a function of the collection semi-angle $\theta_0$ (vanishing at $\theta_{magic}$). A good agreement between the experimental data and the fitting curves (shown in Fig.2) confirms the validity of the expression. As a result, a plot of the pre-factor $F_c$ against the collection semi-angle can give us an accurate value of MA with error normally less than 10%. A set of MA conditions for C 1s absorption determined as a function of the fast electron energy shown in Fig.3a. Our data at 160kV and 200kV were obtained from a JEM-2010F TEM, and the data at 300kV were from a FEI Tecnai F30 TEM. The results reported by Daniels et al[13] are also presented. At 200kV, our result gives MA of 1.2 mrad with 0.1 mrad error, in comparison with Daniels et al's MA[13] of 1.76 mrad and 1.51 mrad obtained from two experiments. We believe that we have the most accurate MA determination so far because it is averaged from a statistical analysis of many independent data points.

Given the importance of MAEELS, a number of theoretical analysis[8,9,12,13] have been performed to explain the observed spectral variation and to calculate the exact value of MA. A recent review[9] shows that all existing conventional theories with the assumptions of the dipole selection rule, single electron excitation and non-relativistic inelastic scattering mechanism should give the same relation:

$$\theta_{magic}=3.97\theta_E \qquad (2)$$

where $\theta_E$ is the characteristic inelastic scattering angle defined as[15]:

$$\theta_E=\Delta E/2E_0 \qquad (3)$$

As shown in Fig.3a, the MA predicted by this non-relativistic (NR) model displays a trend similar with the experimental result with the incident energy, but is quantitatively different. For example, at 200kV, the theoretical MA for the carbon 1s excitation is 2.83 mrad, much larger than the observed MA. This discrepancy is first pointed out by Daniels et al[13] and indicates that there still is a gap in our knowledge even although existing analysis has captured essential physics[9]. A number of other factors may contribute, such as non-dipole transition[15], coherent scattering effect[16], and relativistic effect, but no detailed analysis has been performed.

We have redrawn the experimental data as a fraction of the value predicted by NR

model (Fig.3b) and found out that the deviation seems to increase with the incident energy, thus we suspect that it is due to the relativistic effect. It has often found adequate, in the medium energy range of the fast electrons we have employed, that the relativistic effect can be corrected approximately by modifying the NR model with the substitution of the relativistic corrected value for $\theta_E$[15]:

$$\theta_E^{rel} = \frac{\Delta E}{E_0} \cdot \frac{E_0 + m_0 c^2}{E_0 + 2m_0 c^2} \quad (4)$$

where $m_0$ is the electron rest mass. Physically, this so called 'semi-relativistic' (SR) model only accounts for the over-estimation of the electron wavelength of the fast electrons in NR model. Eq.(4) gives a slightly larger $\theta_E$, hence an even larger theoretical MA value. This is obviously in even worse agreement with the experiment as indicated by the dash doted line in Fig.3a.

To fully account for the relativistic effect, we have to consider the double differential cross-section given by Fano[1] and modified for the dipole approximation used here:

$$\frac{d\sigma}{dE} = \frac{2\pi Z^2 e^4}{mv^2} \left| \frac{i\mathbf{q} \cdot \langle n|\mathbf{r}|0 \rangle}{Q(1+Q/2mc^2)} - \frac{\frac{i\Delta E}{\hbar c}\boldsymbol{\beta}_t \cdot \langle n|\mathbf{r}|0 \rangle}{[Q(1+Q/2mc^2) - \frac{\Delta E^2}{2mc^2}]} \right|^2 \times (1+\frac{Q}{mc^2})dQ \quad (5)$$

where $e$ is the electron charge, $\mathbf{v}$ its velocity, Z the atomic number of the sample, and Q approximately the energy transferred to an unbound electron. The vector $\boldsymbol{\beta}_t$ is defined as the projection of the relativistic vector $\boldsymbol{\beta}$ ($\mathbf{v}/c$) in the direction normal to the vector $\mathbf{q}$, and <n|r|0> is the matrix element of the dipole transition involved. The expression is derived using the so-called Coulomb gauge in which the two transition

amplitudes have clear physical meanings: the first term is called 'longitudinal' due to unretarded static Coulomb interaction that exerts a field parallel to the momentum transfer vector **q**; the second term is called 'transverse' interaction through emission and reabsorption of virtual photons, with a photon field perpendicular to **q**[1].

Up to now, almost all the reported relativistic calculations[15,17-19] of EELS cross-sections have used Fano's relativistic model[20], in which the contribution from the longitudinal and transverse interactions are considered to be added incoherently, i.e. Eq.(5) is expressed as a sum of two transition amplitudes squared separately. Experimentally, Kurata et al[19] has found that this incoherent model is in excellent agreement with the measurement of partial integrated inelastic scattering cross-section for incident energy up to 1MeV and indicated the incoherent relativistic correction introduces at maximum 10% change at 200keV. In our case, Fano's incoherent model predicts that MA for 200keV electrons is about 2% larger than MA predicted by NR model, in contradiction to the experimental value which is 43% smaller.

Close examination of the Fano's incoherent treatment[20] suggests that its validity is only prove for the isotropic case and with transition to all magnetic sublevels included, as it is for Kurata et al's experiment[19]. In general, we found that the cross-term of the two transition amplitude in Eq.(5), when summed over all the possible magnetic quantum states, is proportional to the following:

$$\frac{d\sigma}{dE}\bigg|_{interference} \propto \left[q_x(\beta_t)_x\right]|\langle n|x|0\rangle|^2 + \left[q_y(\beta_t)_y\right]|\langle n|y|0\rangle|^2 + \left[q_z(\beta_t)_z\right]|\langle n|z|0\rangle|^2 \quad (6)$$

where x, y and z are the atom based coordinates. Because $\mathbf{q} \cdot \boldsymbol{\beta}_t = 0$, the condition for the non-observance of this term is satisfied for isotropic systems. By including this interference explicitly in the calculation, we arrive at the fully relativistic (FR) prediction for the MA conditions. When expressed as the correction factor ($MA^{FR}/MA^{NR}$), it is in excellent agreement with experimental result (Fig.3b).

Now we can understand the nature of relativistic correction for medium energy EELS. The pure transverse contribution depending on $\beta^4$ is too small to bring serious correction under 200keV fast electron ($\beta$=0.7). The interference contribution, whose magnitude is proportional to $\beta^2$, dominates the relativistic correction. Its effect on MA is particularly strong because its action on the out of plane excitation (1s->2p($\pi$*)) and in-plane excitation (1s->2p($\sigma$*)) are same in magnitude but opposite in sign.

The existence of the interference effect has been a subject of long-running debate [14]. The apparent success of Fano's popular incoherent model is because the physical systems being considered are mostly isotropic and there is a lack of high quality quantitative experiments in anisotropic systems. As Fano's incoherent theory is at the heart of many cross-sectional calculations[17,21], our conclusion should prompt a re-examination of many such results. Also, our results make it possible to detect the degree of anisotropy using interference effect[22].

Finally, we have expressed the MA conditions in a practically useful form:

$$\theta_{magic} = f(E_0) \cdot 3.97\theta_E \qquad (7)$$

where f(E$_0$) is the relativistic correction factor depending on the incident energy. Following numerical expression can give a good fitting to the fully relativistic results for the incident energy E$_0$ (keV) less than 1MeV with accuracy better than 5%:

$$f(E_0) = (1 + 0.0038 \cdot E_0)^{-1.5} \qquad (8)$$

The energy-loss dependence of MA is solely expressed by θ$_E$. To verify that MA condition is applicable to all anisotropic systems and for all energy losses, we measured the experimental MA for different energy loss using 200keV fast electrons. The result is in good agreement with the Eqs.(7) and (8), as shown in Fig.4.

In summary, we have studied the MA condition experimentally as a function of the energy of fast electrons, as well as a function of the energy loss involved. By using the coherent treatment of relativistic inelastic scattering theory, we obtained an excellent agreement with the experimental results. The surprising large relativistic corrections of the magic angle for medium energy fast electrons gives the first clear evidence for the interference effect between longitudinal and transverse interactions and for the limitation of the Fano's incoherent relativistic model in anisotropic systems. Our result opens a way for the wide use of the MAEELS in which intrinsic electronic structure variation can be measured directly.

**ACKNOWLEDGEMENT**

This research is supported by the Key National Project on Basic Research from Ministry of Science and Technology, the Changjiang Scholar Program of Ministry of Education and the hundred-talent program of the Tsinghua University.    We thank Prof. D.P. Yu of Beijing University for the use of Tecnai TEM facility.

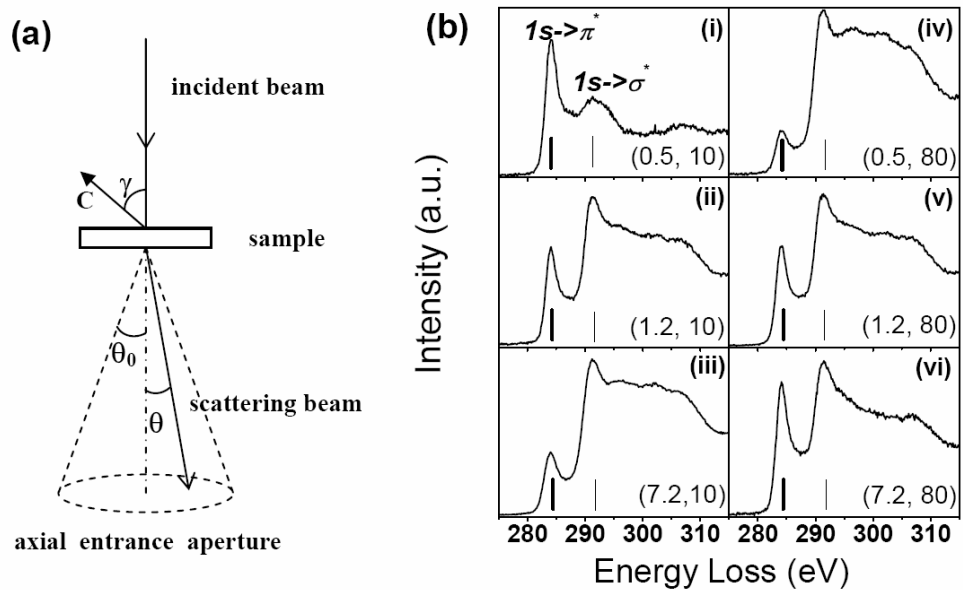

FIG. 1. (a) Experimental geometry for determining magic angle. (b) Selected carbon 1s absorption spectra with the collection semi-angle $\theta_0$ (mrad) and sample orientation $\gamma$ (deg, approximate value) shown in the bracket.

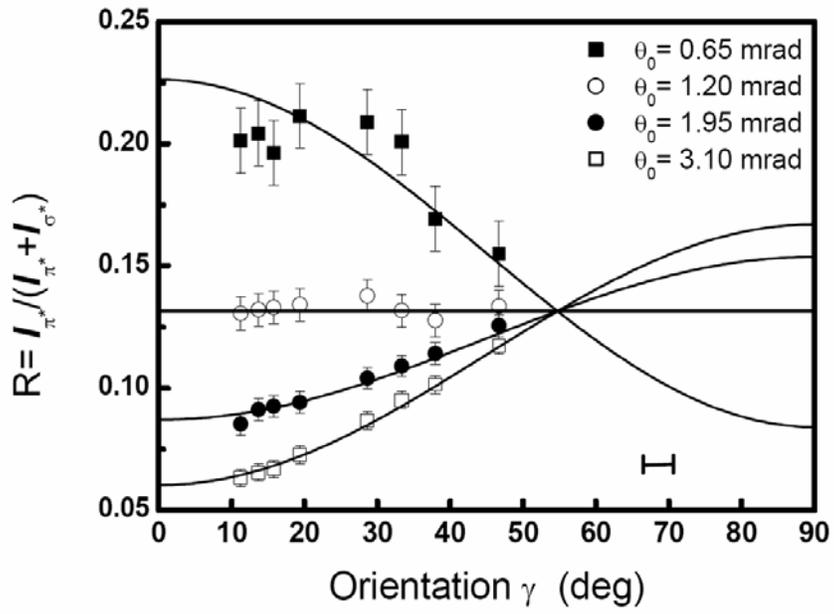

FIG. 2. The plot of normalized ratio $I_{\pi^*}/(I_{\pi^*}+I_{\sigma^*})$ for a series of collection angles as a function of the sample-orientation. The curve fitting (solid lines) based on Eq.(1) shows a good agreement with experimental values.

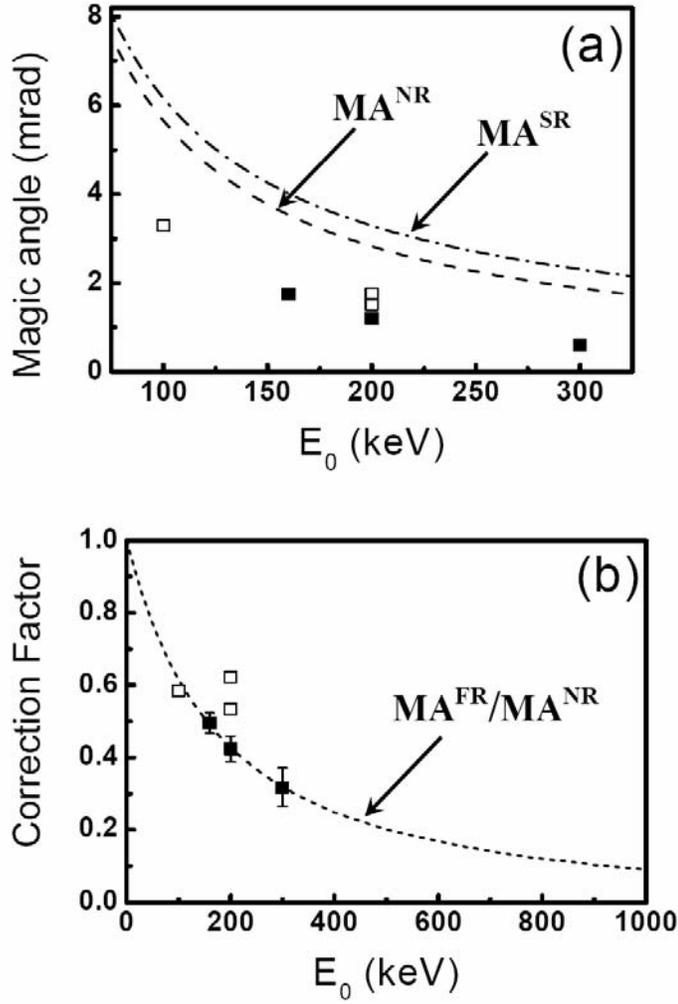

FIG. 3. (a) The comparison between experimentally determined (ours, ■; ref. [13], □) and theoretically predicted MA for carbon 1s absorption with varying incident energy. The dashed line is calculated using non-relativistic (NR) model and dash doted line is using semi-relativistic (SR) model.  (b) Replotting of experimental MA as a fraction of theoretical values predicted by NR model.  For comparison, the dashed line is the correction factor predicted by fully relativistic model (FR).

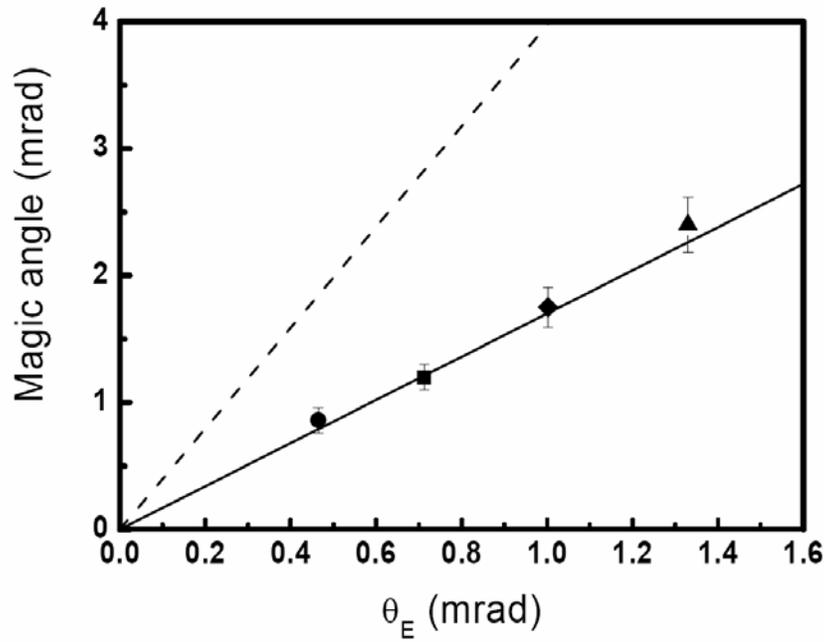

FIG. 4. The MA conditions determined for 1s absorption of boron in BN (●), carbon in Graphite (■), nitride in BN (◆), and oxygen in $B_6O$[23] (▲) for 200keV incident electron. The solid (dashed) line is calculated by the fully (non-) relativistic model, all plotted as a function of the characteristic inelastic angle $\theta_E$ ($\Delta E/2E_0$).